\documentstyle[sprocl,psfig]{article}

\def\to{\rightarrow}
\def\bq{\begin{equation}}
\def\eq{\end{equation}}
\def\ba{\begin{eqnarray}}
\def\ea{\end{eqnarray}}

\def\gsim{\mathrel{\raisebox{-.6ex}{$\stackrel{\textstyle>}{\sim}$}}}
\def\err#1#2{$\stackrel{\scriptstyle +#1}{\scriptstyle -#2}$}
\newcommand{\sla}[1]{/\!\!\!#1}%
\begin{document}

\vspace*{-.6in}

\font\fortssbx=cmssbx10 scaled \magstep1
\hbox to \hsize{
\hbox{\fortssbx University of Wisconsin - Madison}
\hfill\vtop{\hbox{\bf MADPH-98-1081}
            \hbox{October 1998}}}

\vspace{.25in}

\title{NEW INTERACTIONS IN NEUTRAL CURRENT PROCESSES
\footnote{To appear in the 
Proceedings of the 5th International WEIN Symposium: 
A Conference on Physics Beyond the Standard Model (WEIN 98),
Santa Fe, NM, June 14--21, 1998.}
}

\author{DIETER ZEPPENFELD}

\address{Department of Physics, University of Wisconsin,\\ 1150 University 
Avenue, Madison, WI 53706, USA\\ E-mail: dieter@pheno.physics.wisc.edu}

\author{KINGMAN CHEUNG}

\address{Department of Physics, University of California,\\
Davis, CA 95616, USA\\E-mail: cheung@gluon.ucdavis.edu}


\maketitle\abstracts{
Measurements of neutral current processes off the $Z$-peak are sensitive 
probes of new interactions, which may be induced by the exchange of new 
particles such as leptoquarks or extra $Z$'s. This talk reviews the 
phenomenology of extra interactions in four-fermion processes and updates
a global fit to contact terms in the neutral current sector.
}

\section{Introduction}
\label{sec:intro}

Ever since the discovery of weak neutral currents (NC) at 
Gargamelle~\cite{gargamelle} 25 years ago, their study has been a proving 
ground for the Standard Model (SM). During this period, the precision with
which the SM has been confirmed has been improved dramatically, now 
reaching the $0.1\%$ level in $Z$-pole precision experiments.\cite{el98}
Despite these successes of the SM, NC processes remain an excellent
tool to search for, and perhaps discover, signals for new interactions and,
hence, physics beyond the SM. The observation, two years ago at 
HERA~\cite{H1,zeus}, of an apparent excess of events in deep inelastic 
scattering at high $Q^2$ is a case in point, even though 
the excess has not been confirmed by 
subsequent observations.~\cite{hera98}  This possible signal
for new physics has triggered much research, which has lead to a better 
understanding of the experimental constraints on 
new interactions in the NC sector. This talk tries to summarize these
constraints in a model independent way. 
 
Precision experiments on NC interactions can largely be understood
as four-fermion scattering, and among these $\ell\ell qq$ amplitudes are
particularly important. In Section~\ref{sec:NCinSM} we start out by describing
these four fermion NC amplitudes in the SM, and discuss the general form of
deviations due to new interactions. Two such deviations, from very different
sources, are discussed in some detail: extra $Z$ bosons 
are considered in Section~\ref{sec:extraZ}. 
Section~\ref{sec:leptoquarks} deals with leptoquarks, whose low energy
effects can also be described as effective neutral current interactions.

While experiments at the high energy frontier (HERA for leptoquark searches,
the Tevatron for extra $Z$s) can search for direct resonances of these new
particles, most other experiments can be analyzed in a more model-independent
way, via effective four-fermion contact terms. We discuss the general
form of such contact terms in Section~\ref{sec:model.indep.descr}.
The main part of this talk deals with present experimental constraints on 
new NC interactions which can be described in the contact term approximation.
Section~\ref{sec:NCdata} gives a summary of the available data on 
lepton-quark four-fermion interactions, away from the $Z$-peak. These 
data are then used in 
Section~\ref{sec:global.fit} to perform a global fit of $eeqq$ contact
terms. This fit is an update of the one~\cite{bchz97} performed in 1997 and
includes new HERA~\cite{hera98}, LEP2~\cite{lep}, Tevatron~\cite{cdf-dy}, 
and neutrino scattering data~\cite{ccfr,nutev}.
Final conclusions are drawn in Section~\ref{sec:conclusions}.

\section{Neutral Current Exchange within the SM}
\label{sec:NCinSM}
Neutral current four fermion interactions in the Standard Model are due 
to photon and $Z$-exchange. For definiteness, let us consider electron-quark
interactions, as in $e^+e^-\to q\bar q$ annihilation, deep-inelastic scattering
(DIS) experiments, or atomic parity violation (APV). Generalization to
four lepton processes or neutrino-nucleon scattering will be straightforward.

It is convenient to first discuss the amplitudes for the scattering process 
$e_\alpha q_\beta \to e_\alpha q_\beta$, where $\alpha,\beta=L,R$
denote the chirality of the individual fermions. The restriction 
to equal initial and final state chiralities for the electron and quark,
respectively, anticipates chirality conservation in gauge boson interactions.
The scattering amplitude can be decomposed into a reduced amplitude, which
contains the full dynamical information on NC interactions, and wave-function 
factors, which separate out all angular and spin correlation information in 
terms of the external Dirac spinors 
$\psi_i = u(p_i,\sigma_i),\;v(p_i,\sigma_i)$,
\bq
{\cal M}(e_\alpha\; q_\beta \to e_\alpha\; q_\beta)  = 
\overline{\psi_e}\gamma^\mu P_\alpha \psi_e\;\; 
\overline{\psi_q}\gamma_\mu P_\beta \psi_q\;\;
M_{\alpha\beta}^{eq}\; .
\label{eq:fullamp}
\eq
Within the SM, the reduced amplitudes, 
$M_{\alpha\beta}^{eq},\;\alpha,\beta=L,R$, are given by 
\begin{equation}
M_{\alpha\beta}^{eq}(q^2)_{\rm SM} = 
{e^2 Q_e Q_q\over q^2} + {g_Z^2 (T_{e\alpha}^3 - 
s^2_{\rm w}Q_e) (T_{q\beta}^3 - s^2_{\rm w} Q_q) \over q^2 - m_Z^2}\,, 
\label{eq:redamp}
\end{equation}
where $Q_f$ and $T_{f\alpha}^3$ are the charge and third component of 
weak isospin, respectively, 
of the external fermion $f_\alpha$, and the coupling constant factors can 
be written in terms of the weak mixing angle as 
$g_Z = e/(\sin\theta_{\rm w} \, \cos\theta_{\rm w})$, 
$s_{\rm w}=\sin\theta_{\rm w}$. 

A  major advantage of the decomposition of Eq.~(\ref{eq:fullamp}) is that 
the dynamics of crossing related processes is given by the same reduced 
amplitudes, at the appropriate value of the momentum transfer $q$. 
For example, 
\begin{eqnarray}
&& q^2 = \hat s = (E_{\rm c.m.})^2 \hspace{.82in} {\rm for}
\quad e^+e^- \to q\bar q \,,\nonumber\\
&& q^2 = \hat t = -Q^2 = -sxy \hspace{.53in} {\rm for}\quad
 e^\pm p \to ejX\,,\\
&& q^2 = \hat s = sx_1x_2 \hspace{.95in} {\rm for}\quad
 p\bar p\to e^+e^-X\,.
\nonumber
\end{eqnarray}
For $s$-channel processes in the vicinity of the $Z$-pole, one needs to 
replace the $Z$-propagator by its Breit-Wigner form,
\bq
q^2 - m_Z^2 \to \hat s - m_Z^2 +i\;\hat s\;{\Gamma_Z\over m_Z}\; .
\eq
Any new physics contributions to NC processes can be parameterized by adding
an additional term to the reduced amplitude,
\bq\label{eq:ampcontact}
M_{\alpha\beta}^{eq}(q^2) = M_{\alpha\beta}^{eq}(q^2)_{\rm SM} + 
                         \eta_{\alpha\beta}^{eq}(q^2)\; .
\eq
For a large class of new interactions (examples will be given later) the 
new physics contributions $\eta_{\alpha\beta}^{eq}$ vary slowly with $q^2$,
effectively being constant at energies accessible to present experiments.
In this case the $\eta_{\alpha\beta}^{ff'}$ correspond to constant four-fermion
contact interactions (see Section~\ref{sec:model.indep.descr}), and 
Eq.~(\ref{eq:redamp}) relates the sensitivity to new physics of all experiments
probing a given combination of external quarks and leptons, such as 
$ep\to eX$, $p\bar p\to e^+e^-X$, $e^+e^-\to \rm hadrons$ and atomic physics
parity violation experiments.

For a comparison with precision data, electroweak radiative corrections must
be included in the reduced amplitudes discussed above. This is largely achieved
by replacing the coupling constants of the reduced amplitude (\ref{eq:redamp})
by running couplings\cite{hhkm94,Hag.review} 
\ba
e^2 & \to & \bar e^2(q^2)\; ,  \nonumber \\
\sin^2\theta_{\rm w} & \to & {\bar s}^2(q^2) \; \\
g_Z^2 & \to & {\bar g}_Z^2(q^2) \; , \nonumber
\ea
and by adding typically small, process specific vertex and box corrections.

The photon and $Z$ couplings which define the SM amplitudes need to be 
extracted from data, which may contain the effects of both SM and new 
interactions.
A priori, this mingling of SM and new physics effects presents a problem. In 
practice, however, the two decouple. The parameters which define the SM may be 
taken as the fine-structure constant in the Thomson limit, 
$\alpha_{\rm QED}=\bar e^2(0)/4\pi=1/137.036$, the $Z$-mass
and the value of ${\bar s}^2(m_Z^2)$ as determined on the $Z$-pole, the 
top-quark mass as determined at the Tevatron and the Higgs mass obtained 
from fits to $Z$ data and the measured $W$-mass. All these parameters are 
essentially unaffected by the $\eta^{ff'}_{\alpha\beta}$ in 
Eq.~(\ref{eq:ampcontact}). At $q^2\to 0$, photon exchange completely 
dominates the NC amplitudes and the measured fine-structure constant is not 
affected by the new interactions which are due to the exchange of heavy 
quanta. Precision experiments on the $Z$-pole have put very
stringent limits on possible $Z$-$Z'$ mixing effects, and we can safely ignore
them in a phenomenological analysis of $Z'$ effects on other 
NC observables. This 
leaves deviations $\eta^{ff'}_{\alpha\beta}$ from the SM which are 
essentially constant and real in the vicinity of the $Z$-pole and, hence, do 
not interfere with the purely imaginary $Z$-contribution on top of the 
$Z$-resonance. A corollary to this statement
is the fact that $Z$-data are quite insensitive to additional NC contact 
interactions\cite{schrempp} and the constraints on the latter, to be derived 
in Section~\ref{sec:global.fit}, are stringent enough to exclude significant 
effects on the $Z$-pole data. Thus, we can use the fine-structure constant 
and $Z$-pole data to define the SM and then study other NC data to probe 
for possible evidence of new interactions. Because electroweak fits are 
completely dominated by the high 
precision $Z$-data,\cite{chm97}  for our purposes it does not matter whether
one determines the SM parameters from a global fit or from the $Z$ data alone.

\section{Models with extra $Z$ bosons}
\label{sec:extraZ}

Many models of physics beyond the SM involve extensions of the 
$SU(3)\times SU(2)_L \times U(1)$ gauge symmetry, be it grand unified 
theories,\cite{GUTS}  horizontal
symmetries~\cite{horizontal} or extra $W'$ and/or $Z'$ gauge 
bosons.\cite{wprime,zprime}  For our analysis of new interactions in the NC 
sector it is sufficient to consider models with extra $U(1)$ symmetries, i.e.
extensions
\bq
SU(3)\times SU(2)_L \times U(1)_1 \times U(1)_2 \to 
SU(3)\times SU(2)_L \times U(1)_Y \; ,
\eq
where the two extra $U(1)$ factors are spontaneously broken to the hypercharge
symmetry of the SM, leaving behind a massive gauge boson $Z_E$, in addition 
to the still massless hypercharge and $W^3$ gauge fields of the SM. As 
mentioned before, LEP1 and SLC precision data imply that mixing of the 
$Z_E$ with the SM gauge fields is negligible.\cite{chm97}  Let us call $g_E$ 
the gauge coupling of this extra $Z$ boson, and denote the $Z_E$ couplings 
to a fermion $f$ of chirality $\alpha=L,R$ by $g_Eg_\alpha^{Z_Ef}$. The 
exchange of this extra $Z$ boson will alter the
reduced amplitudes of Eq.~(\ref{eq:redamp}) to 
\bq
\label{eq:amp.ZE}
M_{\alpha\beta}^{eq}(q^2) = M_{\alpha\beta}^{eq}(q^2)_{\rm SM} + 
{g_E^2\over q^2-m_{Z_E}^2} g_\alpha^{Z_Ee}g_\beta^{Z_Eq} \; .
\eq
For experiments which operate at $|q^2|<< m_{Z_E}^2$, the $Z_E$ propagator
may be approximated by a constant, and the overall effect of $Z'$ exchange is
an approximately constant addition to the SM reduced amplitude, given by
\bq\label{eq:etaZE}
\eta_{\alpha\beta}^{eq} = 
-{g_E^2\over m_{Z_E}^2} g_\alpha^{Z_Ee}g_\beta^{Z_Eq} \; 
\left(1+{q^2\over m_{Z_E}^2} + \cdots \right)\; .
\eq
The non-observation of an extra $Z$ resonance in Tevatron 
Drell-Yan data\cite{CDF_ZEbound}  suggests
that a $Z_E$, if it exists, must be sufficiently heavy to make the contact term
approximation of Eq.~(\ref{eq:etaZE}) valid for all but the highest energy
experiments, probing momentum transfers in excess of 
$|q^2|\approx 10^5$~GeV$^2$.

The chiral charges $g_\alpha^{Z_Ef} = Q_E$ of the various quark and lepton 
multiplets are model dependent. As an example, let us consider a popular 
class of $Z'$ models, namely extra $U(1)$ symmetries which allow to embed 
the known fermions into 27 dimensional representations of $E_6$. $E_6$ models 
have been studied as candidates for grand unified theories and also as a 
``low-energy'' manifestation of strings.\cite{E6} 

\begin{table}
\caption{Charges of the electroweak fermion multiplets under the $U(1)$ 
factors contained in $E_6$, as given in Eq.~(\protect\ref{eq:E6.decomp}). 
Note that an extra $Z_\psi$ couples purely axially to both quarks and leptons. 
}
\label{table:Q.E6}. 
\begin{tabular}{c|rrr}
        & Y & $\sqrt{24}Q_\chi$ & $\sqrt{72/5}Q_\psi$ \\
\hline
$(\nu,e)_L$  & $-{1\over 2}$ &  3 &  1 \\
$\nu_R$ &        0      &  5 & -1 \\
$e_R$   &       -1      &  1 & -1 \\
$(u,d)_L$    & $ {1\over 6}$ & -1 &  1 \\
$u_R$   & $ {2\over 3}$ &  1 & -1 \\
$d_R$   & $-{1\over 3}$ & -3 & -1 
\end{tabular}
\end{table}

Extra $U(1)$ factors can arise in two steps of the $E_6$ breaking chain
\bq\label{eq:E6.decomp}
E_6 \to SO(10)\times U(1)_\psi \to SU(5)\times U(1)_\chi \times U(1)_\psi\;,
\eq
with two extra $Z$ bosons, $Z_\psi$ and $Z_\chi$, respectively. The $U(1)$ 
charges $Q_\chi$ and $Q_\psi$ are fixed by the embedding of the known fermion 
representations into the 27 of $E_6$, and are given in Table~\ref{table:Q.E6}. 
In these models, in the absence of mixing with the SM $Z$-boson, the lightest 
extra $Z$ boson is a linear combination of the $Z_\psi$ and $Z_\chi$ fields
and consequently the chiral charges $g_\alpha^{Z_Ef}=Q_E$ of (\ref{eq:etaZE}) 
can be written as 
\bq\label{eq:QEmix}
Q_E = Q_\chi \cos\beta_E + Q_\psi \sin\beta_E\; ,
\eq
with the mixing angle $\beta_E$ a free parameter in phenomenological studies,
and the $Q_\chi$ and $Q_\psi$ charges given in Table~\ref{table:Q.E6}.

The most direct observation of a $Z'$ is possible at the Tevatron, as a 
resonance in Drell-Yan $\ell^+\ell^-$ production. The non-observation of 
such a resonance in the 110~fb$^{-1}$ of run~I data has allowed the CDF 
Collaboration to place a lower bound~\cite{CDF_ZEbound} of approximately
\bq
m_{Z_E}\gsim 600 \;{\rm GeV}\qquad {\rm at\; 95\%\; CL}.
\eq
Precise bounds depend somewhat on the value of the mixing angle $\beta_E$, 
and range between 565 and 620~GeV for a discrete set of models considered by
CDF. In addition these mass bounds  assume a $U(1)$ gauge coupling of 
strength similar to the SM hypercharge coupling, $g_E=g_Y=e/\cos\theta_w$. 

\begin{figure}[t]
\centering\leavevmode
\psfig{figure=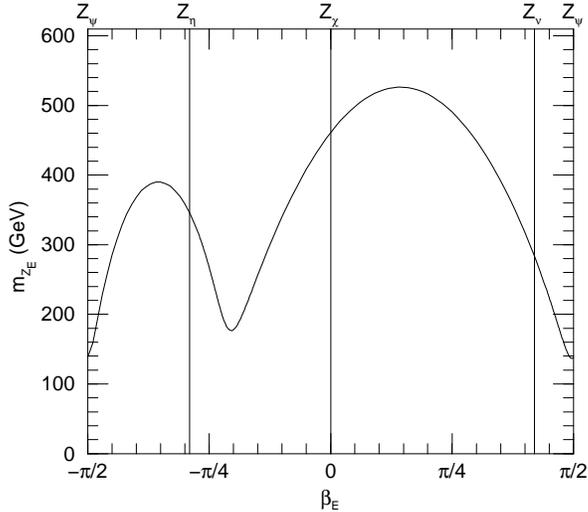,height=2.7in}
\caption{ Lower bound on the mass of an extra $Z$ boson from low energy
NC data, as a function of the $Z_\chi$--$Z_\psi$ mixing angle $\beta_E$ (see 
Eq.~(\protect\ref{eq:QEmix}). No mixing with the SM $Z$ is assumed and 
the $U(1)_E$ gauge coupling is taken of hypercharge-strength, $g_E=g_Y$.
From Ref.~[14].
\label{fig:ZElimits}
}
\end{figure}

For all other present experiments, a $Z_E$ mass in excess of 600~GeV is 
extremely heavy, and fully justifies the contact term approximation of 
Eq.~(\ref{eq:etaZE}). A global analysis, using data from a large set of NC 
experiments, has recently been performed by Cho et al.\cite{chm97}  Results 
are shown in Fig.~\ref{fig:ZElimits}, assuming $g_E=g_Y$. A more general
analysis, allowing an arbitrary $g_E$ coupling and kinetic mixing of the
SM $Z$ and the $Z_E$, has been performed in Ref.~[21], 
demonstrating 
that $Z$-pole data severely limit any mixing effects, while other low energy
data provide the most stringent constraints in the absence of mixing.

Fig.~\ref{fig:ZElimits} shows that the low energy constraints are 
somewhat weaker than the Tevatron bounds from the direct search. 
One must keep in mind, however, that 
the Tevatron sensitivity quickly disappears for $Z_E$ masses closer to the
absolute kinematic limit set by the machine energy of 1.8 TeV.  
The low energy bounds, on the other hand, are on the combination 
$g_E^2/m_{Z_E}^2$ which appears in Eq.~(\ref{eq:etaZE}), and thus also
rule out TeV scale $Z_E$'s with large gauge couplings. Even though
such strongly interacting $U(1)$ gauge bosons are not expected in the context
of GUT models, the different sensitivity of low energy and direct search
experiments to mass and couplings of a $Z_E$ demonstrate their 
complementarity.

\section{Leptoquark models}\label{sec:leptoquarks}

It is fairly obvious that an extra neutral gauge boson $Z_E$ will lead to new
NC interactions. However, very different types of new particles can have 
similar effects at low energies. One prominent example is the exchange of
leptoquarks, scalar or spin-1 bosons whose emission transforms quarks
into leptons and vice versa.\cite{lq,kon91,squark}
We will only consider scalar leptoquarks in the
following. The reason is that within the framework of renormalizable field
theories, elementary vector leptoquarks can only appear as gauge bosons.
Since they must carry both color and electroweak charges they will be the
gauge bosons of a grand unified symmetry and consequently are much too heavy 
to be relevant for measurable effects in NC experiments.\cite{GUTS} 
Composite models might produce TeV scale vector leptoquarks as bound 
states, but since we want to consider 
calculable models only, we disregard this possibility here.

As a first example, motivated by the excitement about a possible $eq$ 
resonance in $ep$ collisions at HERA,\cite{H1}  consider a 
leptoquark $\phi$ which could give such a resonance via \cite{lq}
\bq
e^+ d\to \phi \to e^+d\; .
\eq
The only renormalizable interactions giving this process are the Yukawa
interactions
\bq\label{eq:LQpheno1}
{\cal L}_{LQ}=\lambda \; \overline{d_L}e_R \phi \qquad {\rm with} \quad
Y(\phi)=Y(d_L)-Y(e_R)= {1\over 6}-(-1) = {7\over 6}\;,
\eq
or
\bq
{\cal L}_{LQ}=\lambda \; \overline{d_R}e_L \phi \qquad {\rm with} \quad
Y(\phi)=Y(d_R)-Y(e_L)= -{1\over 3}-\left(-{1\over 2}\right) = {1\over 6}\;.
\eq
Clearly, only one of these terms can be realized since the hypercharge 
assignments, $Y(\phi)$, of the leptoquark are different in the two cases.
In addition, $SU(2)_L$ invariance requires that this leptoquark be a member 
of an $SU(2)_L$ doublet, since the fermion fields have combined weak isospin
1/2.

In the first case, for a $Y(\phi)=7/6$ leptoquark coupling to 
$\overline{d_L}e_R$, the leptoquark contributes to $e^+_Rd_L\to e^+_Rd_L$ 
scattering, and the corresponding amplitude is given by
\bq
{\cal M} = 
-\lambda^2\; \overline{d_L}e_R {1\over q^2-m_\phi^2} \overline{e_R}d_L
= {1\over 2} {\lambda^2\over \hat s-m_\phi^2} \overline{e_R}\gamma^\mu e_R
\overline{d_L}\gamma_\mu d_L \; ,
\eq 
where, in the second step, a Fierz rearrangement has been performed. 
Obviously, $s$-channel leptoquark exchange leads to the same spin structure
as $t$-channel NC exchange, and for momentum transfers well below the 
leptoquark mass the leptoquark contributions leads to an approximately 
constant addition
\bq
\eta_{RL}^{ed}=-{1\over 2} {\lambda^2\over m_\phi^2}
\left(1+{\hat s\over m_\phi^2}+\cdots \right)
\eq
to the reduced amplitude of Eq.~(\ref{eq:redamp}). As a result, leptoquark
effects can be analyzed in a search for new interactions in the NC sector.

This equivalence of leptoquark and NC exchange in low energy experiments is
not limited to the specific model of Eq.~(\ref{eq:LQpheno1}). 
Another example is provided by R-parity violating SUSY models, where an
up-type squark, preferentially a scharm or stop, \cite{squark} acts like a 
leptoquark.
These models are obtained by adding a lepton-number violating but baryon 
number conserving term
\bq
W_{\sla R}= \lambda'_{ijk}L_iQ_jD_k^c
\eq
to the superpotential of the MSSM. Here $L$, $Q$ and $D$ denote the 
superfields describing lepton and quark doublets and the 
righthanded down-quark field, respectively, and $i,j,k$ are generation 
indices. This addition to the superpotential leads to Yukawa couplings 
of squarks and sleptons,
\ba
{\cal L}_{\sla R} = \lambda'_{ijk}\Bigl( &&
\tilde e_L^i\overline{d_R^k}u_L^j+
\tilde u_L^j\overline{d_R^k}e_L^i+ 
\tilde d_R^{k*}\overline{{e_L^i}^c} u_L^j \nonumber \\ 
- && \tilde \nu_L^i\overline{d_R^k}d_L^j -
\tilde d_L^j\overline{d_R^k}\nu_L^i-
\tilde d_R^{k*}\overline{{\nu_L^i}^c} d_L^j 
\Bigr) +{\rm h.c.}\; .
\ea
For $\lambda_{1j1}\ne 0$, the $\tilde u_L^j$ and $\tilde d_L^j$ terms lead
to $ed$ and $\nu d$ scattering via squark exchange. At energies well below
the squark masses these interactions can again effectively be described by a
four-fermion contact interaction,
\ba
{\cal L}_{ed} & = & {(\lambda'_{1j1})^2\over m_{\tilde u_L^j}^2}
\overline{e_L}d_R\overline{d_R}e_L +
{(\lambda'_{1j1})^2\over m_{\tilde d_L^j}^2}
\overline{\nu_L}d_R\overline{d_R}\nu_L \nonumber \\
& = &\left(
-{(\lambda'_{1j1})^2\over 2m_{\tilde u_L^j}^2}\overline{e_L}\gamma^\mu e_L 
-{(\lambda'_{1j1})^2\over 2m_{\tilde d_L^j}^2}\overline{\nu_L}\gamma^\mu\nu_L 
\right) \overline{d_R}\gamma_\mu d_R \;.
\ea
A Fierz rearrangement has again cast the interaction into NC form. Another
interesting feature emerges here. For equal squark masses, $m_{\tilde u_L^j}
=m_{\tilde d_L^j}$, the resulting contact interaction is $SU(2)_L$ symmetric,
\bq
\eta_{LR}^{ed}= -{(\lambda'_{1j1})^2\over 2m_{\tilde u_L^j}^2} =
-{(\lambda'_{1j1})^2\over 2m_{\tilde d_L^j}^2} = \eta_{LR}^{\nu d}\; .
\eq
This relation is expected to be satisfied for squarks of the first two 
generations, but may be violated substantially for stop and sbottom squark 
exchange.

\section{Model independent description}
\label{sec:model.indep.descr}

In the previous two Sections we have seen how very different new physics
contributions can give rise to very similar changes in NC amplitudes, 
provided the mass of the exchanged new particle is well above the 
typical momentum transfer which is accessible experimentally. 
Under this assumption a universal description in terms of NC four-fermion 
contact terms is possible. Let us consider the case of electron-quark
contact interactions in detail. The generalization to other flavors will be 
straightforward.

The most general $eeqq$ current-current contact interaction can be 
parameterized as~\cite{ELP,cashmore,chiap}
\begin{eqnarray}
{\cal L}_{NC} &=& 
\sum_q \Bigl[ \eta_{LL}^{eq}\left(\overline{e_L} \gamma_\mu e_L\right)
\left(\overline{q_L} \gamma^\mu q_L \right) + 
\eta_{RR}^{eq} \left(\overline{e_R}
\gamma_\mu e_R\right) \left( \overline{q_R}\gamma^\mu q_R\right) \nonumber\\
&& \quad {}+ \eta_{LR}^{eq} \left(\overline{e_L} \gamma_\mu e_L\right)
\left(\overline{q_R}\gamma^\mu q_R\right) + 
\eta_{RL}^{eq} \left(\overline{e_R} \gamma_\mu e_R\right)
\left(\overline{q_L} \gamma^\mu q_L \right) \Bigr] \,. 
\label{eq:Leff.pheno}
\end{eqnarray}
The coefficients $\eta_{\alpha\beta}^{eq}$ exactly correspond to the 
extra contributions to reduced amplitudes in Eq.~(\ref{eq:ampcontact}). 
Conventionally they are expressed as 
$\eta_{\alpha\beta}^{eq} = \epsilon 4\pi /\Lambda_{eq}^2$, where
$\epsilon= \pm 1$ allows
for either constructive or destructive interference with the SM $\gamma$ and
$Z$ exchange amplitudes and $\Lambda_{eq}$ is the effective mass scale of the
contact interaction.\cite{ELP} 

Even though contact interactions were originally 
introduced in the context of composite models, the discussion of the previous
Sections shows that they are the tools of choice in a much wider context.
How general, however, is this parameterization of low-energy effects in 
$eeqq$ interactions? Using symmetry arguments we can show that it is general
enough for all practical purposes. From the success of the SM in describing
electroweak precision data\cite{Hag.review} we are confident 
that $SU(2)_L\times U(1)$ gauge invariance is indeed a (spontaneously broken)
symmetry of nature. This implies that the contact term Lagrangian must be
$SU(2)_L\times U(1)$ symmetric, with possible violations only arising from 
mass splittings of the heavy quanta in the same $SU(2)$ multiplet 
(see discussion at the end of Section~\ref{sec:leptoquarks}). 
Symmetry restrictions on the dimension-six effective 
Lagrangian have been analyzed in detail.\cite{wyler} Considering first 
generation fermions only, i.e. limiting the fermion content to the 
fields $L=(\nu_L,e_L)$, $Q=(u_L, d_L)$, $e_R$, $u_R$, and $d_R$, the most
general $SU(2)_L\times U(1)$ gauge invariant contact term Lagrangian of 
energy dimension six can be written as
\begin{eqnarray}
{\cal L}_{SU(2)}&=&
\eta_1  \overline L\gamma^\mu L   \overline Q\gamma_\mu Q 
+
\eta_2  \overline L\gamma^\mu T^aL   \overline Q\gamma_\mu T^a
Q  + \eta_3  \overline L\gamma^\mu L 
 \overline{u_R}\gamma_\mu
u_R \nonumber \\
&& {}+ \eta_4  \overline L\gamma^\mu L 
 \overline{d_R}\gamma_\mu
d_R 
+ \eta_5  \overline{e_R}\gamma^\mu e_R   \overline Q\gamma_\mu
Q  +
\eta_6  \overline{e_R}\gamma^\mu e_R   \overline{u_R}\gamma_\mu
u_R  \nonumber\\
&& {}+ \eta_7  \overline{e_R}\gamma^\mu e_R 
 \overline{d_R}\gamma_\mu d_R  \label{eq:Leff.su2} \\
&& {}+ \Bigl(\eta_8 \overline Le_R \overline{d_R}Q +
\eta_9\overline Le_R \overline Q u_R + \eta_{10} \overline L\sigma^{\mu\nu}
e_R\overline Q\sigma_{\mu\nu}u_R + {\rm h.c.}\Bigr)\; . \nonumber 
\end{eqnarray}
All but the last three are of current-current type and can directly be 
related to the parameters appearing in Eq.~(\ref{eq:Leff.pheno}). The 
$\eta_8$ and $\eta_9$ terms can be written as
\bq
{\cal L}_{8,9} = \eta_8\left(\overline{\nu_L}e_R\overline{d_R}u_L +
                             \overline{e_L}e_R\overline{d_R}d_L \right)
+\eta_9\left(\overline{\nu_L}e_R\overline{d_L}u_R -
                             \overline{e_L}e_R\overline{u_L}u_R \right)
+ {\rm h.c.}\; .
\eq
Both terms induce helicity-non-suppressed $\pi^-\to e^-_R\nu_L$ decay and are
severely constrained by data.\cite{shanker} The same is true for 
the tensor term $\eta_{10}$, which contributes to $\pi^-\to e^-_R\nu_L$ via
a photon loop.\cite{voloshin} Baring cancellations between the three terms,
the experimental constraints on the scalar and tensor coefficients 
$\eta_i = \pm 4\pi/\Lambda_i^2$ can be expressed as~\cite{altCC}
\ba
\Lambda_8,\;\Lambda_9 & \gsim &  500\;{\rm TeV},\;
 \nonumber \\
\Lambda_{10} &\gsim& 90\;{\rm TeV}.\;
\ea
Both bounds are so stringent that no visible signal can be expected in any 
of the NC experiments to be discussed below, and we can safely ignore the
scalar and tensor terms in the following. 

Let us turn back to the current-current type contact terms in the Lagrangian 
of Eq.~(\ref{eq:Leff.su2}). The seven free parameters directly correspond
to the eight different $eeuu$ and $eedd$ coefficients 
$\eta_{\alpha\beta}^{eq}$ in Eq.~(\ref{eq:Leff.pheno}). Thus, $SU(2)$
provides only one constraint among $eeqq$ contact terms, namely
\begin{equation}
\eta^{eu}_{RL}=\eta_5=\eta^{ed}_{RL} \; .
\label{su2releR}
\end{equation}
The main benefit of the $SU(2)$ symmetry is to relate electron and neutrino 
couplings:
\begin{eqnarray}
\eta^{\nu u}_{LL} & = &\eta_1\;+\;{1\over 4}\eta_2 = \eta^{ed}_{LL}\; ,
\nonumber \\
\eta^{\nu d}_{LL} & = &\eta_1\;-\;{1\over 4}\eta_2 = \eta^{eu}_{LL}\; ,
\nonumber \\
\eta^{\nu u}_{LR} & = &\eta_3 = \eta^{eu}_{LR}\; ,
\nonumber \\
\eta^{\nu d}_{LR} & = &\eta_4 = \eta^{ed}_{LR}\; .
\label{eq:su2relnu}
\end{eqnarray}

As is evident from Eq.~(\ref{eq:su2relnu}), $SU(2)$ does not relate the 
$\eta^{eq}_{LL}$ for different quark flavors. The difference 
$\eta^{ed}_{LL}-\eta^{eu}_{LL}=\eta_2/2$ measures the exchange of isospin
triplet quanta between lefthanded leptons and quarks, as indicated by the 
presence of the $SU(2)$ generators $T^a=\sigma^a/2$ in the $\eta_2$ term. 
This term also provides an $e\nu ud$ contact term in CC processes. 
Such contributions, however, are severely limited by lepton-hadron 
universality of weak charged currents,\cite{altCC} within the experimental
verification of unitarity of the CKM matrix. The experimental 
values~\cite{pdg98}
\bq
|V_{ud}^{\rm exp}|=0.9740\pm 0.0010\;,\; 
|V_{us}^{\rm exp}|=0.2196\pm 0.0023\;,\;
|V_{ub}^{\rm exp}|=0.0032\pm 0.0008\;,
\eq
lead to the constraint 
\bq
\left(|V_{ud}|^2+|V_{us}|^2+|V_{ub}|^2\right)
\left(1-{\eta_2\over 4\sqrt{2}G_F}\right)^2 = 0.9969\pm 0.0022\; ,
\eq
when flavor universality of the contact interaction is assumed.
As a result $\eta_2$ must be small, though not necessarily negligible,
\bq\label{eq:eta2CC}
\eta_2 = (0.102\pm 0.073)\;{\rm TeV}^{-2}\;.
\eq

When considering constraints from neutrino scattering experiments in the 
next Section, we will want to invoke $SU(2)$ symmetry and $e$--$\mu$ 
universality in order to restrict 
the number of free parameters. In addition to using the relations of 
Eqs.~(\ref{su2releR}) and (\ref{eq:su2relnu}), we will also impose the CC 
constraint on $\eta_2$ when neutrino data are included in the fits.

In the discussion above we have considered first generation quarks and 
leptons only, because the ``HERA anomaly'' raises particular interest in 
such couplings. In principle, all $\eta$'s carry four independent 
generation indices and may give rise to other flavor conserving and 
flavor changing transitions. The non-observation of intergenerational 
transitions like $K\to\mu e$ puts severe experimental constraints on 
flavor changing couplings.~\cite{wyler} We therefore take a 
phenomenological approach and restrict our discussion to first 
generation contact terms. Only where required by particular 
data (e.g.\ when including $\nu_\mu N$ scattering)
will we assume universality of contact terms between electrons and muons.

\section{Overview of Neutral Current Data}
\label{sec:NCdata}

Many different experiments have been performed to test the NC sector 
of the SM. In the following we are interested in a model independent 
analysis of possible new interactions, in terms of effectively constant
four fermion contact terms. As explained before, these contact terms hardly 
influence the $Z$-pole data, which can therefore be used to extract the 
parameters which define the SM, like $m_Z$, $\sin^2\theta_w$, or the best
fit value for the Higgs mass. In practice, a global fit of all experiments
to the SM parameters gives results very similar to a fit of $Z$-pole data 
only, and therefore we use the ``best fit'' results as given in 1998 by the 
Particle Data Group~\cite{pdg98} to define the SM, against which we compare
off $Z$-pole data in our search for new NC interactions.

The search for new interactions in the leptonic sector, via $ee\ell\ell$
contact terms, is discussed elsewhere in these proceedings.\cite{strovink}
Thus, we concentrate our attention to $eeqq$ contact terms. This analysis 
is an update of the one performed a year ago,\cite{bchz97} and we follow the
basic procedure developed there. Relevant constraints
arise from a variety of experiments:
\begin{enumerate}
\item
 atomic parity violation (APV)~\cite{apv}
\item
 polarized lepton nucleon scattering~\cite{eD,mainz89,bates90}
\item
 DIS at HERA\cite{H1,zeus,hera98}
\item
 Drell-Yan production of lepton pairs at the Tevatron~\cite{cdf-dy}
\item
 $e^+e^-\to$ hadrons at LEP2~\cite{lep} and
\item
 $\nu_\mu$--nucleon scattering.\cite{ccfr,nutev}
\end{enumerate}
For the first five data sets we can perform a completely model-independent 
analysis, in terms of the eight parameters $\eta_{\alpha\beta}^{eu}$ and
$\eta_{\alpha\beta}^{ed}$. 
In order to include neutrino-nucleon scattering data, we must
assume $e$--$\mu$ universality and $SU(2)$ symmetry of the contact terms,
as given by Eq.~(\ref{eq:su2relnu}). We now discuss these various data 
sets in turn. Additional details can be found in Refs.~[6,14].

\subsection{Atomic Parity Violation}

Parity violation in the SM is due to weak gauge boson exchange, with
vector-axial-vector ($VA$) and axial-vector-vector ($AV$) terms contributing.
Atomic physics parity violation experiments measure the weak charge $Q_W$ 
of heavy atoms, which is given by~\cite{langacker}
\begin{equation}
Q_W = -2 \biggr[ C_{1u} (2Z+N) + C_{1d} (Z+2N) \biggr] \;,
\end{equation}
where $Z$ and $N$ are the number of protons and neutrons respectively in the
nucleus of the atom. Here the $C_{1q}$ are the coefficients describing 
$A_eV_q$ couplings in a contact term description 
of standard model $eeqq$ NC couplings. The new physics contact terms 
of Eq.~(\ref{eq:Leff.pheno}) lead to a shift of these couplings,
\bq\label{eq:deltaC1q}
\Delta C_{1q} = {1\over 2\sqrt{2}G_F}
\left(\eta_{RL}^{eq}+\eta_{RR}^{eq}-\eta_{LL}^{eq}-\eta_{LR}^{eq}\right) \;.
\eq

Recently a very precise measurement was made of the parity violating transition
between the 6S and 7S states of $^{133}_{\phantom055}$Cs with the use of a 
spin-polarized atomic beam.~\cite{apv} Accounting for a slight improvement
in the atomic theory calculation,\cite{dzuba} the resulting $Q_W$ is given 
as~\cite{el98}
\begin{equation}
Q_W^{\rm exp} = -72.41\pm 0.25 \pm 0.80 \;, \label{weakcharge}
\end{equation}
where the first uncertainty is experimental and the second is theoretical.
This needs to be compared with the value for Cs predicted by the SM,
including radiative corrections,\cite{el98}
\begin{equation}
Q_W^{\rm SM}=-73.10 \pm 0.04 \,.
\end{equation}
This theoretical value agrees with the measurement within errors, leading to
the constraint
\bq
\Delta C_{1u} + 1.122 \Delta C_{1d} = -0.0018\pm 0.0022 \;.
\eq
Some chirality combinations of $LL,RR,LR,RL$ give zero
contributions to $\Delta C_{1q}$ and thus satisfy the experimental $Q_W$
constraint trivially. Such possibilities include (i) $LL=RR=LR=RL \;(VV)$,
(ii) $LL=RR=-LR=-RL\;(AA)$, (iii) $LL=-LR, RL=-RR$ (an SU(12) 
symmetry~\cite{nelson}), (iv) $LR=RL, LL=RR=0$ (a minimal choice used 
in fitting the HERA data \cite{ours}).

\subsection{Polarization Asymmetries in Electron-Nucleus Scattering 
Experiments}

Experiments on polarized electron-nucleus scattering measure the 
polarization asymmetries 
\bq
A = {d\sigma(e^-_R N) - d\sigma(e^-_L N) \over 
     d\sigma(e^-_R N) + d\sigma(e^-_L N) }
\eq
and are sensitive to both $A_eV_q$ and $V_eA_q$ combinations of 
electron and quark currents. The former are parameterized by the $C_{1q}$
discussed above. $V_eA_q$ combinations are conventionally parameterized by 
coefficients $C_{2q}$ in a contact term description of $eeqq$ NC 
couplings.\cite{pdg98} The extra contributions to NC interactions in 
(\ref{eq:Leff.pheno}) then lead to shifts in $C_{1q}$ 
(see (\ref{eq:deltaC1q})) and in $C_{2q}$,
\bq\label{eq:deltaC2q}
\Delta C_{2q} = {1\over 2\sqrt{2}G_F}
\left(-\eta_{RL}^{eq}+\eta_{RR}^{eq}-\eta_{LL}^{eq}+\eta_{LR}^{eq}\right) \;.
\eq

The most precise data on polarized lepton-nucleon scattering come from 
three experiments. Comparing with the SM 
expectations, the results of these experiments can be summarized as 
follows:~\cite{chm97,bchz97} the SLAC $e$-D scattering 
experiment~\cite{eD} gives
\ba
\Delta( 2C_{1u}-C_{1d}) & = & -0.22 \pm 0.26 \,, \qquad
\Delta( 2C_{2u}-C_{2d}) = 0.77 \pm 1.23 \,,  \nonumber \\
\rho_{\rm corr} & = & -0.975\,,
\ea
where $\rho_{\rm corr}$ is the correlation. 
The Mainz $e$-Be scattering experiment\cite{mainz89} yields the constraint
\bq
\Delta C_{1u} - 0.24 \Delta C_{1d} + 0.80 \Delta C_{2u} - 0.74 \Delta C_{2d}
= -0.024 \pm 0.070\;.
\eq
Finally, the Bates $e$-C scattering experiment\cite{bates90} implies
\bq
\Delta C_{1u} + \Delta C_{1d} = -0.015 \pm 0.033\;.
\eq

\subsection{$e^+ p$ scattering at HERA}

In early 1997, the H1~\cite{H1} and ZEUS~\cite{zeus} experiments at DESY 
reported an excess of DIS events at large $Q^2$. The possibility of a 
significant deviation from the SM has triggered much of the recent research 
on new interactions in the NC sector. Since the original publications, the 
amount of data collected by the two HERA groups has roughly been doubled. 
Our analysis is based on the HERA data as presented at the 1998 spring and 
summer conferences~\cite{hera98} which correspond to a combined 
integrated luminosity of over 80~pb$^{-1}$. The input to our fit is listed 
in Table~\ref{table:hera}, together with the SM expectation and its 
theoretical error.

\begin{table}[t]
\caption[]{
\label{table:hera}
\small The measured and expected number of events as a function of 
$Q^2_{\rm min}$ at HERA.}
\medskip
\centering
\begin{tabular}{|ccl||ccl|}
\hline
\multicolumn{3}{|c||}{ZEUS (${\cal L}=46.60\, {\rm pb}^{-1}$) }& 
\multicolumn{3}{c|}{H1 (${\cal L}=37.04\, {\rm pb}^{-1}$) } \\
\hline
\underline{$Q^2_{\rm min} \,({\rm GeV}^2)$} & \underline{$N_{\rm obs}$}
 & \underline{$N_{exp}$} & 
\underline{$Q^2_{\rm min} \,({\rm GeV}^2)$} & \underline{$N_{\rm obs}$} &
\underline{$N_{exp}$} \\
2500 & 1817 & 1792$\pm93$ &  2500 & 1297 & 1276$\pm98$ \\ 
5000 & 440  & 396$\pm24$ &  5000  & 322  & 336$\pm29.6$ \\
10000 & 66  & 60$\pm4$ &  10000  & 51   & 55.0$\pm6.42$ \\
15000 & 20  &  17$\pm2$ & 15000  & 22   & 14.8$\pm2.13$ \\
35000 & 2   & 0.29$\pm0.02$ & 20000 & 10   & 4.39$\pm0.73$ \\
      &     &      & 25000 & 6    & 1.58$\pm0.29$ \\
\hline
\end{tabular}
\end{table}

Neutral current deep inelastic scattering occurs via the subprocess 
$eq\to eq$. In terms of the reduced amplitudes
$M^{eq}_{\alpha \beta} (\hat t) \;(\alpha,\beta=L,R)$ of 
Eq.~(\ref{eq:ampcontact}) the spin- and color-averaged amplitude-squared 
for $e^+q\to e^+q$ is given by 
\ba
&&\overline{\sum} |{\cal M}(e^+ q\to e^+ q)|^2 = \nonumber \\
&& \biggr( |M^{eq}_{LL}(\hat t)|^2 + |M^{eq}_{RR}(\hat t)|^2 \biggr) \hat u^2 +
 \biggr( |M^{eq}_{LR}(\hat t)|^2 + |M^{eq}_{RL}(\hat t)|^2 \biggr) \hat s^2
   \;, \label{ampl^2}
\ea
where $\hat s = sx$ is bounded by the HERA center of mass energy, 
$s\simeq(300{\rm GeV})^2$, $\hat t = -Q^2$, and $\hat u = -\hat s + Q^2$.
Because additional contact terms are most important at large $Q^2$, which
favors small $|\hat u|$, $e^+q$ scattering at HERA is most sensitive to
$\eta_{RL}$ and $\eta_{LR}$ contact terms. $\eta_{LL}$ and $\eta_{RR}$
contributions are enhanced by the $\hat s^2$ factor in $e^+$-antiquark
collisions, but here the smaller antiquark distributions at large $x$ 
lead to a loss in sensitivity. Similarly, because $u(x,Q^2)>d(x,Q^2)$,
and because the SM couplings of the up-quark are larger in DIS than those
of the down-quark, leading to larger interference contributions of the 
$\eta^{eu}$, the HERA experiments are most sensitive to deviations in 
the $M^{eu}_{RL}$ and $M^{eu}_{LR}$ reduced amplitudes.

\subsection{Drell-Yan Cross Section at the Tevatron}

The production of Drell-Yan lepton pairs at the Tevatron, via the subprocess
$q\bar q\to e^+e^-$, probes $eeqq$ 
contact terms at the highest accessible energies. The double differential 
cross section, versus the lepton pair invariant mass, $\sqrt{\hat s}$, and 
rapidity, $y$, is given by
\begin{equation}
\frac{d^2\sigma}{d\hat s dy} =
K\, \frac{\hat s}{144\pi s} \sum_q q(x_1) \bar q(x_2)
\sum_{\alpha,\beta=L,R} |M_{\alpha\beta}^{eq}(\hat s)|^2\; ,
\end{equation}
where $q(x_1)$ and $\bar q(x_2)$ are parton distributions evaluated at
$Q^2=\hat s$, $x_{1,2} = e^{\pm y}\sqrt{\hat s/s}$, and 
the reduced amplitudes $M_{\alpha\beta}^{eq}(\hat s)$ are given by
Eqs.~(\ref{eq:redamp},\ref{eq:ampcontact}). 
$K$ is the QCD $K$-factor for Drell-Yan production.

Since our previous analysis, 
CDF has published the observed number of events in bins of invariant mass
of the lepton pair, as given in Table~\ref{table:cdfdy}. We use these 
data for our fit, together with the SM expectation provided 
by CDF.\cite{cdf-dy} 
The SM expectation is normalized to the number of events seen at the 
$Z$ peak, which effectively fixes the $K$-factor.
Since the data are consistent with $e$-$\mu$ universality, we 
use both the electron and muon data. The Tevatron data extend out to very
large lepton invariant masses, where they exclude large deviations
in any of the squared reduced amplitudes. This eliminates the possibility
of large cancellations between different flavor and helicity combinations
of the $\eta_{\alpha\beta}^{eq}$, and is crucial for reducing the 
correlations in the global fit.

\begin{table}[t]
\caption[]{ \label{table:cdfdy}
\small The electron and muon samples of Drell-Yan production in 
CDF~\protect\cite{cdf-dy} together with the SM expectation. }
\medskip
\centering
\begin{tabular}{|c|cc||cc|}
\hline
& 
\multicolumn{2}{c||}{$e^+ e^-$} & \multicolumn{2}{c|}{$\mu^+ \mu^-$} \\
\hline
\underline{$M_{\ell\ell}$} & \underline{$N_{\rm obs}$}
 & \underline{$N_{\rm exp}$} & \underline{$N_{\rm obs}$}
 & \underline{$N_{\rm exp}$}  \\
50--150  &  2581 & 2581 & 2533 & 2533 \\ 
150--200 &  8    & 10.8 &   9  & 9.7  \\
200--250 &  5    & 3.5  &   4  & 3.2  \\
250--300 &  2    & 1.4  &   2  & 1.3  \\
300--400 &  1    & 0.97 &   1  &  0.94 \\
400--500 &  1    & 0.25 &   0  &  0.27 \\
500--600 &  0    & 0.069 &  0  & 0.087 \\
\hline
\end{tabular}
\end{table}

\subsection{$e^+e^-\to q\bar q$ at LEP2}

The same reduced amplitudes as in Drell-Yan production, however with 
somewhat different weighting, are measured in $e^+e^-\to$~hadrons.
At leading order in the electroweak interactions, the total hadronic cross
section for $e^+e^-\to q\bar q$, summed over all flavors $q=u,d,s,c,b$, is
given by
\begin{equation}
\sigma_{\rm had} = K\, \sum_q \frac{s}{16\pi} \biggr[
|M_{LL}^{eq}(s)|^2 + |M_{RR}^{eq}(s)|^2 + |M_{LR}^{eq}(s)|^2 +
|M_{RL}^{eq}(s)|^2 \biggr] \,,
\label{sig_had}
\end{equation}
where $K=1+\alpha_s/\pi+1.409(\alpha_s/\pi)^2-12.77(\alpha_s/\pi)^3$
is the QCD $K$ factor.\cite{gorishny91} 

Since charge or flavor identification of light quark jets is problematic,
we only consider the LEP2 results on the total hadronic cross section in our
analysis, summed over five quark flavors. Contact interactions are included 
in $e^+e^-\to u\bar u$ and $e^+e^-\to d\bar d$ amplitudes only. 

The LEP collaborations have presented measurements of the total hadronic
cross section at center of mass energies between $\sqrt{s}=130$ and 
189 GeV.\cite{lep}  The data used
for our global fit are summarized in Table~\ref{table:LEP2}.

\begin{table}[t]
\caption[]{
\label{table:LEP2}
\small Total hadronic cross sections $\sigma_{\rm had}$ measured
by the LEP collaborations and SM expectations.\protect\cite{lep} }
\medskip
\centering
\begin{tabular}{|ccc|}
\hline
$\sqrt{s}$ (GeV) &  $\sigma_{\rm had}$  & $\sigma_{\rm SM}$ \\
\hline
\multicolumn{3}{|c|}{ALEPH} \\
\hline
130 & 79.5 $\pm$ 4.14 &  77.16 \\
136 & 64.5 $\pm$ 3.85 &  62.52 \\
183 & 23.6 $\pm$ 0.73 & 23.05 \\
\hline
\multicolumn{3}{|c|}{DELPHI}\\
\hline
130.2 & 82.2 $\pm$ 5.2 &  83.1 \\
136.2 & 65.9 $\pm$ 4.7 &  67.0 \\
161.3 & 40.2 $\pm$ 2.1 &  34.8 \\ 
172.1 & 30.6 $\pm$ 2.0 &  28.9 \\
\hline
\multicolumn{3}{|c|}{L3}\\
\hline
130.3 & 81.8 $\pm$ 6.4 &  78 \\
136.3 & 70.5 $\pm$ 6.2 &  63 \\
140.2 & 67   $\pm$ 47  &  56 \\
161.3 & 37.3 $\pm$ 2.2 &  34.9 \\ 
170.3 & 39.5 $\pm$ 7.5 &  29.8 \\
172.3 & 28.2 $\pm$ 2.2 &  28.9 \\
\hline
\multicolumn{3}{|c|}{OPAL} \\
\hline
130.25 & 64.3 $\pm$ 5.1 &  77.6 \\
136.22 & 63.8 $\pm$ 5.2 &  62.9 \\
161.34 & 35.5 $\pm$ 2.2 &  33.7 \\ 
172.12 & 27.0 $\pm$ 1.9 &  27.6 \\
183    & 23.7 $\pm$ 0.81&  24.3 \\
189    & 21.8 $\pm$ 0.89&  22.3 \\
\hline
\end{tabular}
\end{table}

\begin{table}[t]
\caption[]{\label{table:fit}
\small
The best estimate of the $\eta_{\alpha\beta}^{eq}$ parameters when various
data sets are added successively.  In the last column, when the $\nu N$
data and the CC universality constraint of Eq.~(\protect\ref{eq:eta2CC})
are included, the $\eta_{L\beta}^{\nu q}$ are given in terms of
$\eta_{L\beta}^{eq}$ by the SU(2) relations of Eq.~(\protect\ref{eq:su2relnu}) 
and we assume $\eta_{RL}^{eu}=\eta_{RL}^{ed}$.
}
\medskip
\centering
\begin{tabular}{|l|c|c|c|c|c|}
\hline
 & HERA only  & +APV+eN  & +DY      &  +LEP     &    +$\nu N$ \\
\hline
$\eta_{LL}^{eu}$ & 2.04\err{3.97}{5.26}  
  & 2.24\err{2.29}{3.63}  & 0.22\err{0.67}{0.57}
  & -0.08\err{0.58}{0.35}  & 0.02\err{0.22}{0.23} \\
$\eta_{LR}^{eu}$ & -4.30\err{4.30}{0.78} 
  & -2.77\err{3.20}{1.70}  & 0.60\err{0.51}{0.66}
 & 0.77\err{0.35}{0.62}  & 0.41\err{0.22}{0.28} \\
$\eta_{RL}^{eu}$ & -1.74\err{3.75}{2.60} 
  & -3.53\err{2.91}{0.90}  &  0.00\err{0.72}{0.72}
  & 0.00\err{0.76}{0.74}  & 0.49\err{0.31}{0.39} \\
$\eta_{RR}^{eu}$ & 2.62\err{4.29}{5.35}  
 & 2.23\err{1.77}{3.41} & 0.04\err{0.66}{0.62}
 & -0.14\err{0.66}{0.53} & -0.25\err{0.47}{0.33} \\
$\eta_{LL}^{ed}$ & -1.72\err{7.87}{6.76} 
 & -2.23\err{5.62}{4.54}  & 0.25\err{1.65}{1.71}
 & 0.04\err{0.65}{0.80}  & 0.07\err{0.23}{0.24} \\
$\eta_{LR}^{ed}$ & -0.01\err{4.85}{4.48} 
 & -0.95\err{3.76}{3.47}  & 1.65\err{1.39}{2.79}
 & 0.04\err{1.29}{1.47}  & 0.50\err{0.57}{0.61} \\
$\eta_{RL}^{ed}$ & -1.87\err{4.87}{4.37} 
 & -0.95\err{3.87}{3.23}  & 1.98\err{1.30}{2.74}
 & 0.36\err{1.25}{1.44}  & $=\eta_{RL}^{eu}$ \\
$\eta_{RR}^{ed}$ & -2.26\err{7.85}{7.24} 
 & -1.60\err{5.59}{4.85}  & 0.55\err{1.60}{1.73}
 & 0.40\err{0.76}{0.91}  & 0.20\err{0.55}{0.64} \\
\hline
\hline
HERA    & 7.57 & 7.86  & 12.10 & 12.80  & 13.06  \\
APV+eN  &      & 0.48  & 0.48  &  0.48  & 1.07  \\
DY      &      &       & 4.40  &  4.36  & 4.26 \\
LEP     &      &       &       &  21.45 & 21.54 \\
$\nu N$ &      &       &       &        &  0.10   \\
\hline
\hline
Total $\chi^2$ & 7.57 & 8.34   & 16.98 & 39.08 & 40.04  \\
\hline
SM $\chi^2$ & 17.27 &  19.79 & 24.08 & 45.91 & 49.22 \\
\hline
SM d.o.f.   & 11    &  16     & 28     & 47     & 50 \\
\hline
\end{tabular}
\end{table}

\subsection{Neutrino-Nucleon DIS Experiments}

Deep inelastic scattering experiments with neutrino and anti-neutrino
beams have provided important tests for the SM since the early
80's.\cite{haidt,global} Assuming $e$-$\mu$ universality and SU(2)$_L$ 
invariance, we can use $\nu_\mu N$ DIS data to constrain the lepton-quark
contact interactions of Eq.~(\ref{eq:Leff.su2}). 

While older measurements are available (see e.g. the discussion in Ref.~[2]),
the most precise data come from recent CCFR and NuTeV
measurements. The CCFR collaboration has obtained a model-independent 
constraint on the effective $\nu\nu qq$ couplings from a measurement of the
NC to CC cross section ratio:~\cite{ccfr}
\ba
\kappa & = & 1.6981\left(g_L^u\right)^2 + 
1.8813\left(g_L^d\right)^2 + 1.0697\left(g_R^u\right)^2 + 
1.2261\left(g_R^d\right)^2 \nonumber \\ 
       & = & 0.5820 \pm 0.0041 \;, \label{ccfr_kappa}
\ea
which should be compared to the SM 
prediction~\cite{pdg98} $\kappa=\kappa_{\rm SM} = 0.5830 \pm 0.0005$.
Here $g_L^q,g_R^q$ (called $\epsilon_L(q),\epsilon_R(q)$ in Ref.~[32])
are the coefficients in the neutrino-quark effective Lagrangian describing 
their NC interactions. 

A second combination of these couplings appears in the Paschos-Wolfen\-stein
parameter and has recently been measured by the NuTeV collaboration. 
They find~\cite{nutev}
\begin{equation}
0.8587\left(g_L^u\right)^2 + 0.8828\left(g_L^d\right)^2 
-1.1657\left(g_R^u\right)^2 - 1.2288\left(g_R^d\right)^2 
= 0.2277 \pm 0.0022 \;, 
\label{nutev_R-}
\end{equation}
compared to a SM value of $0.2302\pm0.0003$.

The presence of new interactions in the NC sector, as parameterized by
Eq.~(\ref{eq:Leff.su2}), leads to shifts in the $g_{L,R}^q$ given by
\begin{equation}
\Delta g_L^q =  - \frac{1}{2\sqrt{2}G_F} \eta_{LL}^{\nu q}\;,\qquad
\Delta g_R^q =  - \frac{1}{2\sqrt{2}G_F} \eta_{LR}^{\nu q} \;, 
\label{eq:Delta.gLRnuq}
\end{equation}
as compared to the SM expectations given in Ref.~[32]. 
The $\eta_{\alpha\beta}^{\nu q}$ are related to the corresponding
$eeqq$ contact terms by Eq.~(\ref{eq:su2relnu}). When using these SU(2)
relations to constrain physics beyond the SM, we also impose the concomitant
CC constraint of Eq.~(\ref{eq:eta2CC}).

\begin{table}[t]
\caption[]{\label{table:fit.single}
\small
The best estimate on $\eta_{\alpha\beta}^{eq}$ and the 95\% CL limits on the
compositeness scale $\Lambda_{\alpha\beta}^{eq}$,
where $\eta_{\alpha\beta}^{eq}=4\pi\epsilon/(
\Lambda_{\alpha\beta\epsilon}^{eq})^2$.
When one of the $\eta$'s is considered the others are set to zero.
SU(2) relations are assumed and $\nu N$ data are included.
}
\medskip
\centering
\begin{tabular}{|ccc|cc|}
\hline
& & & \multicolumn{2}{c|}{95\% CL Limits} \\
Chirality  ($q$) &  $\eta\;\;$ (TeV$^{-2}$)  &  $\chi^2_{\rm min}$ &
$\Lambda_+$ (TeV) & $\Lambda_-$ (TeV) \\
\hline
\hline
LL($u$)  & $-0.036 \pm 0.029$ &47.61 &  18.4  & 12.1 \\
LR($u$)  & $0.094  \pm 0.067$ &47.26 &  7.8   & 12.2 \\
RL($u$)  & $-0.028\pm 0.035$  &48.56  &  15.5 & 11.9 \\
RR($u$)  & $-0.079\pm 0.070$  &47.93  &  11.5 & 8.0 \\
LL($d$)  & $0.057 \pm 0.029$  &45.51  &  10.9 & 19.8 \\
LR($d$)  & $0.041 \pm 0.064$  &48.81  &  9.0  & 11.1\\
RR($d$)  & $-0.046\pm 0.065$  &48.71  &  11.2 & 8.8 \\
\hline
\end{tabular}
\end{table}

\section{Discussion}
\label{sec:global.fit}

Table~\ref{table:fit} summarizes the results of our new global fit, at various 
steps of including additional data. In the first four columns the eight
phenomenological parameters $\eta^{eq}_{\alpha\beta}$ are treated as free.
$e$-$\mu$ universality is assumed when 
considering Tevatron Drell-Yan data. Only in the last
column, when adding constraints from $\nu N$ scattering, do we assume the 
SU(2) relations of Eqs.~(\ref{su2releR},\ref{eq:su2relnu}), and, thus, this
column represents a seven parameter fit. Note that here the CC constraint of
Eq.~(\ref{eq:eta2CC}) is also included in the fit.

The $\chi^2$ per degree of freedom ($\chi^2_{\rm cont.}$/d.o.f.=0.931) 
of the contact interactions is very close to that of the SM 
($\chi^2_{\rm SM}$/d.o.f.=0.984) for the last column in Table~\ref{table:fit},
and both are excellent fits. Thus, no signal for new interactions in the NC
sector is found in the present data.
As is apparent from the jump in $\chi^2$ for the HERA data alone, 
from 7.86 to 12.1 when going from column 2 to 3, there
is some clash between the still persistent excess of high $Q^2$ events 
at HERA with the Tevatron Drell-Yan data. However, one should note
that this is mainly due to the pre-1997 HERA data. The effect would be 
minimal if 1997 data only had been considered. 

The errors listed for the fit parameters in Table~\ref{table:fit} are fairly 
large and caused by strong correlations between the $\eta^{eq}_{\alpha\beta}$. 
The correlations hide the stringency of constraints, in particular from
the APV and $\nu N$ scattering experiments. Table~\ref{table:fit.single}
gives a complementary view by fitting the data with a single nonzero
$\eta_{\alpha\beta}^{eq}=4\pi\epsilon/(\Lambda_{\alpha\beta\epsilon}^{eq})^2$ 
at a time. Also included are the $95\%$~CL limits on the corresponding 
$\Lambda_{\pm }$. Note that the
$\pm$ definition is from the sign of the terms in the Lagrangian.  So,
whether the ``$+$'' sign will interfere constructively or destructively
with the SM amplitude depends on the sign of the SM amplitude.
We find that parity violating new interactions, which
contain $A_eV_q$ current-current couplings, must have an intrinsic scale
$\Lambda>10$~TeV.

\begin{table}[t]
\caption[]{\label{table:noAPV.ud}
\small
The best estimate on $\eta^{eq}$ for the minimal setting, $VV,AA$, and
SU(12), and the corresponding 95\% CL limits on the compositeness scale 
$\Lambda$, where $\eta=4\pi\epsilon/(\Lambda_{\epsilon})^2$. Apart from 
the $\eta$-combination specified, all the others are set to zero. Here 
we do not use SU(2) relations and, hence, do not include the $\nu N$ data.
}
\medskip
\centering
\begin{tabular}{|ccc|cc|}
\hline
& & & \multicolumn{2}{c|}{95\% CL Limits} \\
Chirality  ($q$) &  $\eta\;\;$ (TeV$^{-2}$) & $\chi^2_{\rm min}$ &
$\Lambda_+$ (TeV)  & $\Lambda_-$ (TeV) \\
\hline
\hline
$\eta_{LR}^{eu}=\eta_{RL}^{eu}$  & 0.42\err{0.23}{0.25}&43.31 & 4.0 & 6.6 \\
$\eta_{LR}^{ed}=\eta_{RL}^{ed}$  & $-0.67$\err{0.47}{0.32}&44.34&2.7 & 3.3 \\
\hline
$\eta_{VV}^{eu}$   & $-0.0092$\err{0.078}{0.076}&45.90 &  9.1 & 9.1 \\
$\eta_{VV}^{ed}$   & 0.11\err{0.16}{0.19} &45.53&  5.8 & 5.0 \\
\hline
$\eta_{AA}^{eu}$  & $-0.17$\err{0.11}{0.099} &43.45& 9.0  &  6.2 \\
$\eta_{AA}^{ed}$  & 0.11 \err{0.12}{0.14}   &45.24& 6.3  &  7.3 \\
\hline
$\eta_{LL}^{eu}=-\eta_{LR}^{eu}$  & $-0.28$\err{0.18}{0.16}&43.70 &6.3& 4.8 \\
$\eta_{RL}^{eu}=-\eta_{RR}^{eu}$  & 0.33\err{0.18}{0.20}&43.41 & 4.5 & 5.5 \\
$\eta_{LL}^{ed}=-\eta_{LR}^{ed}$  & 0.19\err{0.22}{0.25}&45.29  & 4.8 & 4.9 \\
$\eta_{RL}^{ed}=-\eta_{RR}^{ed}$  & $-0.23$\err{0.28}{0.24}&45.20& 3.8 & 4.5 \\
\hline							    
\end{tabular}						    
\end{table}

\begin{table}[th]
\caption[]{\label{table:noAPV.q}
\small
Same as the last Table but with a further condition: $\eta^{eu}=\eta^{ed}$.
Here $q=u=d$ and SU(2) constraints and the $\nu N$ data are included.
}
\medskip
\centering
\begin{tabular}{|ccc|cc|}
\hline
& & & \multicolumn{2}{c|}{95\% CL Limits} \\
Chirality  ($q$) &  $\eta\;\;$ (TeV$^{-2}$) & $\chi^2_{\rm min}$ &
 $\Lambda_+$ (TeV)  & $\Lambda_-$ (TeV) \\
\hline
\hline
$\eta_{LR}^{eq}=\eta_{RL}^{eq}$  & 0.40 \err{0.19}{0.22}&46.15 & 4.3  & 6.0 \\
\hline
$\eta_{VV}^{eq}$   & $0.0092$\err{0.12}{0.11}&49.21  &  6.9  & 8.1 \\
\hline
$\eta_{AA}^{eq}$  & $-0.29$\err{0.13}{0.12}&44.77  & 8.3   & 5.2 \\
\hline
$\eta_{LL}^{eq}=-\eta_{LR}^{eq}$  & $-0.33$\err{0.17}{0.16}&45.65 &7.6 & 4.7\\
$\eta_{RL}^{eq}=-\eta_{RR}^{eq}$  & 0.50\err{0.20}{0.25} &46.25&4.0 & 4.0\\
\hline
\end{tabular}
\end{table}

For new interactions in the NC sector which do not affect APV observables,
the constraints can be considerably weaker. This is demonstrated in 
Tables~\ref{table:noAPV.ud} and \ref{table:noAPV.q} where several APV 
blind combinations of $\eta_{\alpha\beta}^{eq}$ are considered. Note that in 
this comparison we cannot impose $SU(2)\times U(1)$ symmetry without relating 
$eeuu$ and $eedd$ couplings. Hence the fits of Table~\ref{table:noAPV.ud}, 
which consider couplings to either up- or down-quarks,
do not include $\nu N$ scattering data. While $VV$ and $AA$ interactions
still require scales $\Lambda$ in excess of 5 to 9~TeV, some fairly large
contact terms are allowed for special combinations, the largest being
$\eta_{LR}^{ed}=\eta_{RL}^{ed}$, for which scales as small as 3~TeV are still 
allowed.

Much of the recent interest in new interactions in the NC sector has been
motivated by the apparent excess of high $Q^2$ events in the pre-1997 
HERA data~\cite{H1,zeus}. Comparing the results of a HERA only fit with
the results of the global fit in Table~\ref{table:fit} it may appear that
the severe constraints by other experiments, at LEP2 or the Tevatron
for example, exclude the observation of signals for new interactions
at a machine like HERA. Is this conclusion justified?

\begin{figure}[t]
\centering\leavevmode
\psfig{figure=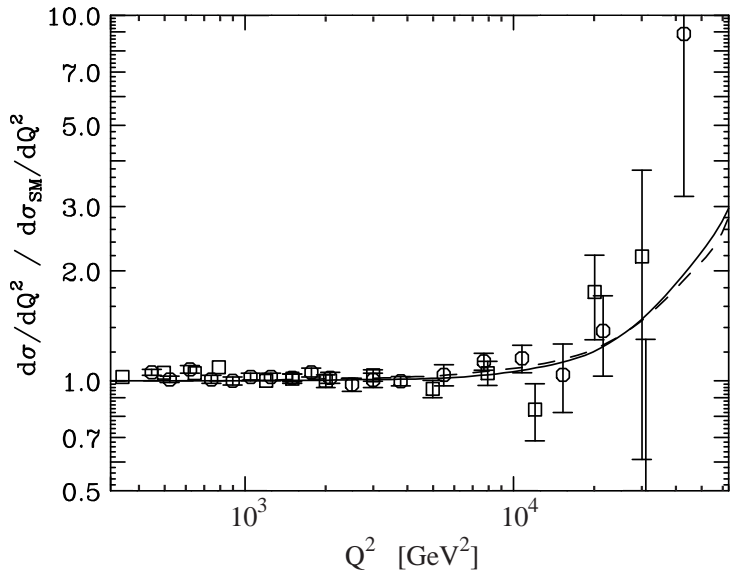,height=3.0in}
\vspace*{-0.1in}
\caption{
Ratio of the neutral current cross section $d\sigma^{NC}/dQ^2$ to the 
SM expectation. Shown are H1 data~\protect\cite{H1dsdQ} (squares) and 
ZEUS data~\protect\cite{ZEUSdsdQ} (circles). Errors are statistical only. 
Superimposed are the expectations for
two contact term choices: the best global fit result in the last column
of Table~\protect\ref{table:fit} (solid line), and the choice 
$\eta_{LR}^{eu}=\eta_{RL}^{eu}=0.42$~TeV$^{-2}$ (dashed line) (see 
Table~\protect\ref{table:noAPV.ud}). 
\label{fig:hera}
}
\vspace*{-0.2in}
\end{figure}

The most straightforward signature for new interactions at HERA is an
enhancement in the 
differential DIS cross section, $d\sigma^{NC}/dQ^2$. In Fig.~\ref{fig:hera}
the present measurements~\cite{H1dsdQ,ZEUSdsdQ} 
are compared with the SM expectation.
Also included in this plot is the ratio of non-standard over SM cross 
sections expected for the best fit result, in the last column of 
Table~\ref{table:fit} (solid line), and the analogous cross section ratio for
the case with the smallest $\chi^2$ in Table~\ref{table:noAPV.ud}, 
$\eta_{LR}^{eu}=\eta_{RL}^{eu}=0.42$~TeV$^{-2}$ (dashed line).
The best fits clearly favor an interpretation of the excess high $Q^2$ events
as a statistical fluctuation. However, it is also obvious that sizable and 
significant signals for new NC interactions remain possible with higher 
statistics at HERA.

\section{Conclusions}
\label{sec:conclusions}

After 25 years of neutral current experiments, the Standard Model has been
confirmed to amazing precision. Nevertheless, ongoing and future experiments
may well discover new physics in NC processes. There are many possible
sources of such new interactions. History might repeat and the exchange of
an extra $Z$ boson could lead to new neutral quark and lepton currents. 
However, also other phenomena, which are not a priori of current-current
type, may give rise to effective neutral current interactions, leptoquark
exchange being one example. 

More general new interactions between quarks and leptons are possible, but
our knowledge of chirality conservation in $eeqq$ interactions and the 
apparent $SU(2)_L\times U(1)$ symmetry of nature relegates 
such more general scalar and tensor interactions to very high scales.
This makes it quite possible that new quanta
in the 1 to 10 TeV range will first be observed indirectly, as new
neutral current interactions.

The range of competitive experiments in this field is truly remarkable, 
ranging from atomic physics parity violation measurements, over DIS 
experiments, to $e^+e^-$ annihilation at LEP, and Drell-Yan pair production
at the Tevatron. None of these experiments dominates the field. APV and 
$\nu_\mu N$ scattering experiments have similar sensitivities to
parity violating observables, and the sensitivity of LEP and the Tevatron,
while slightly lower for specific couplings, is needed to exclude 
cancellations of different couplings in the other experiments. Only
future experiments can tell whether nature has such new interactions in store
for us or whether the SM is a perfect model for NC data at the much 
higher energies yet to be explored.

\vspace*{-0.1in}

\section*{Acknowledgments}
This research was supported in part by the University of Wisconsin Research
Committee with funds granted by the Wisconsin Alumni Research Foundation and
by the Davis Institute for High Energy Physics, and 
in part by the U.~S.~Department of Energy under Contract
Nos. DE-FG02-95ER40896 and DE-FG03-91ER40674.

\end{document}